\documentclass[apj]{emulateapj}

\begin{document}

\submitted{Accepted for publication in the Astropysical Journal Letters on August 7, 2014}

\shorttitle{``Old'' Pulsars in Supernova Remnants}
\shortauthors{Bogdanov et al.}

\title{Constraining the Evolutionary Fate of Central Compact Objects: \\ ``Old'' Radio Pulsars in Supernova Remnants}

\author{Slavko Bogdanov\altaffilmark{1}, C.-Y. Ng\altaffilmark{2}, Victoria M.~Kaspi\altaffilmark{3}}

\altaffiltext{1}{Columbia Astrophysics Laboratory, Columbia University, 550 West 120th Street, New York, NY 10027, USA; slavko@astro.columbia.edu}

\altaffiltext{2}{Department of Physics, The University of Hong Kong, Pokfulam Road, Hong Kong}

\altaffiltext{3}{Department of Physics, McGill University, 3600 University Street, Montreal, QC H3A 2T8, Canada}

\begin{abstract}  
Central compact objects (CCOs) constitute a population of radio-quiet,
slowly-spinning ($\ge$100 ms) young neutron stars with anomalously
high thermal X-ray luminosities.  Their spin-down properties imply
weak dipole magnetic fields ($\sim$$10^{10-11}$ G) and characteristic
ages much greater than the ages of their host supernova
remnants (SNRs). However, CCOs may possess strong
``hidden'' internal magnetic fields that may re-emerge on timescales
$\gtrsim$10 kyr, with the neutron star possibly activating as a radio
pulsar in the process.  This suggests that the immediate descendants
of CCOs may be masquerading as slowly spinning ``old'' radio pulsars.
We present an X-ray survey of all ordinary radio pulsars within 6 kpc
that are positionally coincident with Galactic SNRs in
order to test the possible connection between the supposedly old, but
possibly very young pulsars, and the SNRs. None of the targets exhibit
anomalously high thermal X-ray luminosity, suggesting that they are
genuine old ordinary pulsars unrelated to the superposed SNRs. This
 implies that CCOs are either latent radio pulsars that activate long after
 their SNRs dissipate or they remain permanently radio-quiet.
 The true descendants of CCOs remain at large.
\end{abstract}


\section{INTRODUCTION}
In the absence of other means, the age of a rotation-powered pulsar is
estimated from its characteristic age, defined as $\tau_c\equiv P/2
\dot{P}$, where $P$ is the pulsar period and $\dot{P}$ is the spindown
rate.  Although not exact, this value is generally a rough
approximation of the true age of a pulsar, provided that the birth
spin period was much shorter than the current period.  However,
\citet{Halpern10} and \citet{Got13a} have recently measured unusually
low spindown rates of three CCOs, an enigmatic
group of young, slowly-spinning ($P\gtrsim100$ ms), radio-quiet but
X-ray bright neutron stars at the centers of supernova remnants, which
indicate unusually low magnetic fields ($\sim$$10^{10-11}$ G) and
characteristic ages orders of magnitude greater than the host SNR
ages.  A profound implication of this result is that the
characteristic age grossly overestimates the true age of these neutron
stars. By extension, many supposedly old radio pulsars
($\tau_c\ge10^5$ yr) may, in fact, be relatively young (a few to tens
of kyr).  This may occur, for instance, if CCOs provide an alternative
formation channel that ``injects'' long period ($>$$100$ ms), low
spin-down rate neutron stars into the pulsar population. Such objects
may be masquerading as typical pulsars, in which case the conventional
picture of pulsar evolution in the $P$$-$$\dot{P}$ diagram may require
substantial revisions \citep[e.g.,][]{Fau06}.

The spin properties of the three CCOs with detected periodicities (PSR
J1852+0040 in Kes 79, PSR 0821--4300 in Puppis A, and 1E1207.4--5209
in PKS 1209--51/52) imply relatively weak surface magnetic fields of
$\sim$$10^{10}$ G.  However, the highly non-uniform surface
temperature distribution deduced from the thermal pulsations
\citep{Got10,Bog14} suggest the presence of much stronger subsurface
fields. These ``hidden'' strong fields are expected to re-emerge on
time-scales 1--100 kyr \citep[see, e.g.,][]{Ho11,Vig12}, depending on
the submergence conditions, transforming CCOs into neutron stars with
stronger external fields ($\sim$$10^{12}$ G) at a later evolutionary
stage and possibly activating as radio pulsars in the process.

%
%   FIGURE 1
%
\begin{figure}[!t]
\begin{center}
\includegraphics[width=0.45\textwidth]{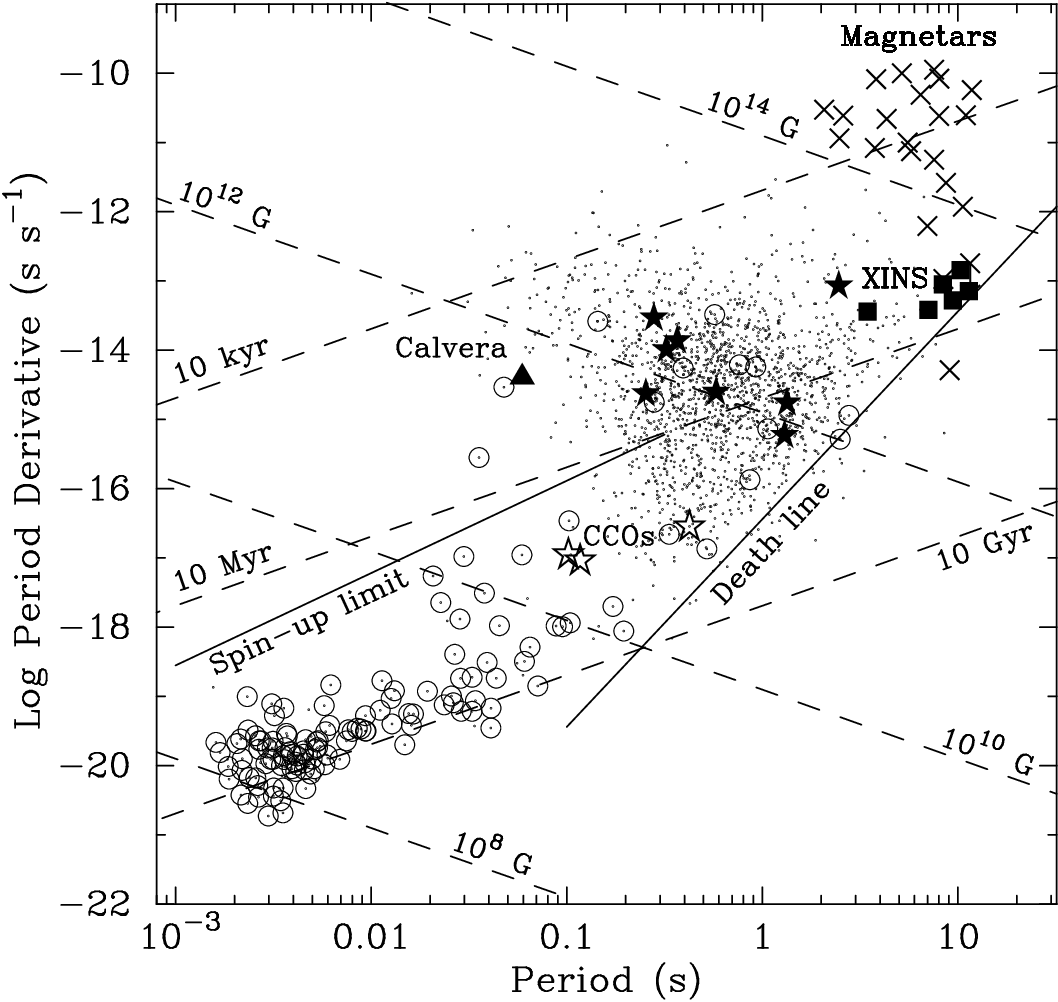}
\end{center}
\caption{The $P-\dot{P}$ diagram of neutron stars.~The open stars mark
  the three CCOs with measured spin-down rates
  \citep{Halpern10,Got13a}, while the solid stars show the eight
  pulsars considered in this paper. The radio- and $\gamma$-ray quiet
  X-ray pulsar Calvera is shown with the solid triangle. The dots show
  all radio pulsars from the ATNF pulsar catalog with the binary
  pulsars marked with a circle.  The crosses show the population of
  magnetars, and the solid squares show X-ray dim isolated neutron
  stars (XINS). The solid line shows the theoretical death line,
  beyond which pulsars cease to generate radio emission, and the
  spin-up limit, while the dashed lines show tracks of constant age
  and magnetic fields.}
\end{figure}

In this evolutionary scenario, some ordinary middle-aged and old radio
pulsars with $\sim$$10^{12}$ G may in fact be relatively young CCOs,
especially those situated within or very near the boundaries of
SNRs. These remnants may actually be the pulsar birth sites even
though such associations may be dismissed as chance superpositions
based solely on the pulsar's high value of $\tau_c$.  One possible
manifestation of the youth of these pulsars should be unusually hot
thermal emission from the pulsar, resulting in relatively high X-ray
luminosity ($L_X\ge10^{33}$ erg s$^{-1}$) comparable to or in excess
of the pulsar spindown luminosity, as seen in CCOs.  To test this
possibility we investigate the X-ray emission from the eight
middle-aged/old radio pulsars within 6 kpc that are positionally
coincident with Galactic SNRs.  This study has important implications
for understanding the birth, evolution, and properties of the Galactic
population of neutron stars, especially in light of recent
discoveries.

\begin{deluxetable*}{lllccccccc}
\tabletypesize{\small} 
\tablecolumns{11} 
\tablewidth{0pc}
\tablecaption{{\small Parameters of the eight pulsars targeted in this study. }}
% \centering
%  \hline 
%  \hline
\tablehead{
\colhead{Pulsar} & \colhead{$\alpha$ (J2000)} & \colhead{$\delta$ (J2000)}  & \colhead{$P$} & \colhead{$D$\tablenotemark{a}} &  \colhead{DM} &  \colhead{$\tau_c$} & \colhead{$\dot{E}$} &  \colhead{$B_{s}$} & Ref. \\
  \colhead{ }   & \colhead{(h m s)} &   \colhead{($^{\circ}$ $'$ $''$)}    &   \colhead{(s)}  &  \colhead{(kpc)}   & \colhead{(pc cm$^{-3}$)}  &  \colhead{(Myr)}   &  \colhead{($10^{33}$ erg s$^{-1}$)} & \colhead{($10^{12}$ G)} & \colhead{}}
\startdata
  B0905--51   & 09~07~15.90 & $-$51~57~59.2   & 0.25 &  2.7    &  104   & $2.2$     & $4.4$  &  0.69 & 1,2 \\
  B1703--40   & 17~07~21.72 & $-$40~53~56.1  & 0.58  &  5.1    &  360   & $4.8$     & $0.39$  &  1.1 & 3,4 \\
  B1736--29   & 17~39~34.27 & $-$29~03~03.5  & 0.32  &  3.2    &  139   & $0.65$    & $9.2$  & 1.6  & 5,6 \\
  B1742--30   & 17~45~56.30 & $-$30~40~23.5  & 0.37  &  2.1    &  88    & $0.55$    & $8.5$  &   2.0 & 7,8 \\
  J1808--2701 & 18~08~13.23 &  $-$27~01~21  & 2.46   &  2.4    &  95    & $0.59$    & $0.17$  & 12.9 & 9 \\
  B1822--14   & 18~25~02.92 & $-$14~46~52.6 & 0.28   &  5.0    &  357   & $0.20$    & $41$  &  2.6 & 5,6 \\ 
  J1901+0254  & 19~01~15.67 &  $+$02~54~41 & 1.30    &  3.5    &  185   & $45$      & $0.0082$ & 0.78 &  10 \\
  B1919+21    & 19~21~44.81 & $+$21~53~02.2  & 1.33  &  1.0    &  12    & $16$      & $0.022$ & 1.4 & 11,8
\enddata
\tablenotetext{a}{Distance derived from the pulsar dispersion measure and the NE2001 model for the Galactic
distribution of free electrons \citep{Cor02}.}
%\end{data}
\tablerefs{ (1)~Manchester et al.~1978; (2)~Siegman et al.~1993; (3)~Johnston et al.~1992; (4)~Wang et al.~2001; (5)~Clifton \& Lyne~1986; (6)~Hobbs et al.~2004b; (7)~Komesaroff et al.~1973; (8)~Zou et al.~2005; (9)~Lorimer et al.~2006; (10)~Hobbs et al.2004a; (11)~Hewish et al.~1968.}
\label{table:1}
\end{deluxetable*}

\section{Sample Selection}
To identify radio pulsars that fall within or just outside of the
boundaries of SNRs, we have cross-correlated the catalogue of Galactic
Supernova remnants\footnote{See
  http://www.mrao.cam.ac.uk/surveys/snrs/} \citep{Green09} and the
ATNF pulsar catalogue\footnote{Available at
  http://www.atnf.csiro.au/research/pulsar/psrcat/} \citep{Man05}.
After filtering out the well-established associations, we have
narrowed down the list to a volume-complete sample of eight known old
pulsars within 6 kpc.  For reference, the chance spatial coincidence
probability of a pulsar falling within the confines of any given SNR
in the Galactic plane is $\sim$5\%. Although
this value is by no means negligible, there is still a strong
possibility that at least some of the pulsars and SNRs are truly
associated.  We have searched the HEASARC archive for serendipitous
pointings towards these objects. Five pulsar positions fall within
existing \textit{Chandra} and/or \textit{XMM-Newton} images.  In order
to obtain a volume-complete sample within 6 kpc, we have targeted the
remaining three objects, PSR J1808--2701, J1901+0254, and B1919+21
with \textit{XMM-Newton}.  The basic parameters of all eight objects
are summarized in Table 1 and their locations in the $P-\dot{P}$
diagram are shown in Figure 1.  All eight objects have spin periods
$P>0.25$ s and are representative of the population of ordinary radio
pulsars.  Their distances were estimated based on the dispersion
measure (DM) combined with the NE2001 model for the Galactic
distribution of free electrons \citep{Cor02}. Among the eight SNRs,
only G272.2--3.2, coincident with PSR B0905--51, has a published
distance estimate of $1.8^{+1.4}_{0.8}$ kpc \citep{Harrus01}, which is
consistent with the DM-derived pulsar distance of 2.7 kpc.

\begin{deluxetable}{llccc}
\tabletypesize{\small} 
\tablecolumns{5} 
\tablewidth{0pc}
\tablecaption{Summary of X-ray Observations.}
% \centering
%  \hline 
%  \hline
\tablehead{
\colhead{Pulsar} & \colhead{Telescope/} & \colhead{Observation}  & \colhead{Epoch} & \colhead{Exp.} \\ 
  \colhead{ }   & \colhead{Instrument} &   \colhead{ID}    &  \colhead{ } & \colhead{(ks)}} 
\startdata
  B0905--51   & \textit{XMM}/EPIC  & 011293101   & 2001-12-10  &  37.4    \\
              & \textit{CXO}/ACIS     & 10572       & 2008-08-27  &  22.8   \\
              & \textit{CXO}/ACIS     & 9147        & 2008-08-26  &  41.6   \\
  B1703--40   & \textit{XMM}/EPIC  & 0144080101  & 2002-09-27  &  16.8    \\
              & \textit{XMM}/EPIC  & 0406580101  & 2006-08-25  &  26.4    \\
  B1736--29   & \textit{CXO}/ACIS     & 8678        & 2008-05-18  &  ~2.2   \\
              & \textit{CXO}/ACIS     & 8679        & 2008-05-18  &  ~2.2   \\
  B1742--30   & \textit{XMM}/EPIC  & 0103261301  & 2001-03-21  &  ~7.6    \\
              & \textit{CXO}/ACIS     & 8747        & 2008-05-15  &  ~2.2   \\
  B1822--14   & \textit{CXO}/ACIS     & 4600        & 2004-07-09  &  11.0   \\
              & \textit{CXO}/ACIS     & 5341        & 2004-07-11  &  18.0   \\
\hline
  J1808--2701 & \textit{XMM}/EPIC  & 0692210101	& 2012-09-15  &  21.9   \\
  J1901+0254  & \textit{XMM}/EPIC  & 0692210301  & 2013-03-29  &  16.9   \\
  B1919+21    &  \textit{XMM}/EPIC & 0670940101  & 2012-03-20 &  16.9 
\enddata
\label{table:3}
\end{deluxetable}
 
\section{Data Reduction and Analysis}
Table 2 lists the X-ray data used in this analysis.
We performed the re-processing, reduction, and analysis of the
\textit{Chandra} data with CIAO\footnote{Chandra
  Interactive Analysis of Observations, available at
  \url{http://cxc.harvard.edu/ciao/}} 4.6 and the corresponding
calibration products CALDB version 4.5.9 \citep{Fruscione06}.  For the
\textit{XMM-Newton} data, we used the Science Analysis Software
(SAS\footnote{The \textit{XMM-Newton} SAS is developed and maintained
  by the Science Operations Centre at the European Space Astronomy
  Centre and the Survey Science Centre at the University of
  Leicester.}) version {\tt xmmsas\_20130501\_1901-13.0.0}.  The
\textit{XMM-Newton} events lists were cleaned by applying the
recommended flag, pattern, and pulse invariant filters, and screening
for instances of severe background flares.

We extracted events within 2$\arcsec$ of the radio pulsar position for
the \textit{Chandra} data and 40$''$ for the \textit{XMM-Newton}
MOS1/2 data. Data from \textit{XMM-Newton} EPIC PN was not used as we
found it to be more severely affected by strong background flares,
resulting in much shorter exposures relative to the MOS data. The
analysis was restricted to the 0.3--3 keV band, where most of the
thermal radiation from pulsars is typically detected with these
telescopes.  The background was taken from a source-free region in
the image near the pulsar position on the same detector chip.

\begin{deluxetable*}{lcccccccc}
\tabletypesize{\small} 
\tablecolumns{9} 
\tablewidth{0pc}
\tablecaption{{\small Parameters of the eight supernova remnant -- pulsar superpositions. }}
% \centering
%  \hline 
%  \hline
\tablehead{
\colhead{}           & \colhead{Supernova} & \colhead{Remnant}  & \colhead{Angular} & \colhead{$\tau_c$} & \colhead{Kin.~age\tablenotemark{b}} &  \colhead{$N_H$\tablenotemark{c}}  & \colhead{$L_X$\tablenotemark{d}} & Refs. \\
  \colhead{Pulsar}   & \colhead{remnant}    &   \colhead{diameter}      &  \colhead{offset\tablenotemark{a}}      &  \colhead{(Myr)}   &  \colhead{(kyr)} & \colhead{($10^{21}$ cm$^{-2}$)} &  \colhead{($10^{31}$ erg s$^{-1}$)} & \colhead{ }}
\startdata
 B0905--51   &  G272.2--3.2 &  $\sim$$15'$      & $11.1'$        & $2.2$   & 22  & 3.1  & 2.3  & 1,2,3 \\
 B1703--40   &  G345.7--0.2 &  $6'$             & $1.03'$       & $4.8$    & 4   & 11  & $<$22   & 4,5 \\
 B1736--29   &  G359.1+0.9  &  $12'\times11'$   & $8.0'$        & $0.65$   & 19  & 4.2  & $<$4.6  & 6,7,8 \\
 B1742--30   &  G358.5--0.9 &  $\sim$$17'$      & $3.4'$        & $0.55$   & 28  & 2.6  & $<$2.6    & 9,7 \\
 B1822--14   &  G16.8--1.1  &   $30'\times24'$  & $4.4'$        & $0.20$   & 10  & 11  & 6.9   & 10 \\ 
 J1808--2701 &  G4.2--3.5   & $28'$             & $10.5'$       & $0.59$ & 19 &  2.9 & $<$2.1   & 11,12\\
 J1901+0254  &  G36.6--0.7  & $25'$             & $11.0'$       & $45$   & 27 & 5.6    & $<$13   & 13,14 \\
 B1919+21    &  G55.7+3.4   & $23'$             & $11.1'$       & $16$   & 11 & 0.36  & $<$0.14   & 15
\enddata
\tablenotetext{a}{Angular separation between the pulsar and the estimated center of the SNR.}
\tablenotetext{b}{Estimated kinematic age (i.e.~the time required for the pulsar to traverse the distance from the SNR center to its current position) based on the pulsar distance and transverse velocity, where available. Where no proper motion information is available, a velocity of 380 km s$^{-1}$ is assumed.}
\tablenotetext{c}{Hydrogen column density along the line of sight computed based on the empirical relation $N_{\rm H}(10^{20}~{\rm cm}^{-2}) \simeq 0.30{\rm DM(pc cm^{-3})}$ found by He et al.~(2013).}
\tablenotetext{d}{Estimated bolometric luminosity or 2$\sigma$ upper limit assuming thermal emission with $kT=0.2$ keV.} 
\tablerefs{(1)~Greiner et al.~1994; (2)~Duncan et al.~1997; (3)~Harrus et al.~2001; (4)~Whiteoak \& Green 1996; (5)~Green et al.~1997; (6)~Gray 1994a; (7)~Roy \& Bhatnagar 2006; (8)~Law et al.~2008; (9)~Gray 1994b; (10)~Reich et al.~1986; (11)~Reich et al.~1988; (12)~Reich et al.~1990; (13)~F\"urst et al.~1987; (14)~Kassim 1992; (15)~Goss et al.~1977.}
\label{table:2}
\end{deluxetable*}

\section{Results}
 The characteristic ages of the pulsars considered here range from 0.2
 to 45 Myr.  If these pulsars were in fact born in the centers of the
 SNRs, their characteristic ages do not correspond to their true age,
 as is the case for CCOs. In this scenario, we can estimate their
 kinematic ages by calculating the travel time from the remnant center
 to the current position.  For PSR B1919+21, a proper motion
 measurement is available and implies a velocity of 190 km s$^{-1}$ at
 the DM-derived distance of 1 kpc \citep{Zou05}. It is
 interesting to note that the proper motion vector of the pulsar
 points in the direction away from the center of the remnant. For PSR
 B1822--14 (aka PSR J1825--1446), \citet{Mol12} report a proper
 motion-derived transverse velocity of 690 km s$^{-1}$ at a distance
 of 5 kpc. Although the direction of the pulsar's proper motion is
 generally consistent with the pulsar moving away from the inner
 regions of the remnant, due to the irregular morphology of G16.8--1.1
 and contamination in the radio from the bright source RCW 164, the
 exact center of the remnant is difficult to determine. As a result,
 we can only crudely estimate an angular separation of $\sim$4$'$
 between the pulsar and the inner region of the SNR.

For the remaining pulsars, we assume the mean velocity of pulsars in
the Galaxy of 380 km s$^{-1}$ derived by \citet{Fau06} and the DM
distances.  Based on this, the putative pulsar-remnant associations in
Table 2 imply ages in the range $4-28$ kyr, orders of magnitude
smaller than their $\tau_c$.

If the pulsars are indeed that young, their surfaces should be at
substantially higher temperatures (and thus luminosities) compared to
$10^{5-6}$ yr-old pulsars.  Assuming this emission is powered by
residual cooling, if we scale the luminosity of the CCO in Kesteven
79, PSR J1852+0040, based on the neutron star cooling curves given in
\citet[][see in particular, their Figure 11]{Vig13}, we estimate that
for ages $\lesssim10^5$ yr it would have a blackbody temperature of
$kT \gtrsim 0.37$ keV with a bolometric luminosity 
$\gtrsim$$10^{33}$ ergs s$^{-1}$.  On the other hand, typical old pulsars
have $L_X\sim10^{-5}-10^{-3}\dot{E}$ suggesting that the eight pulsars
in Table 2 would have $L_X\lesssim10^{32}$ ergs s$^{-1}$ if they are
solely rotation-powered \citep[see, e.g.,][]{Kaspi06}. Hence, an
observed $L_X$ much greater than this would be strong evidence for a
young neutron star.

The results of the analysis for the eight pulsars are summarized in
Table 3.  Six pulsars are not detected in X-rays: PSRs B1703--40,
B1742--30, B1736--29, J1808--2701, J1901+0254, and B1919+21
implying bolometric luminosities $L_X\lesssim 10^{32}$ ergs s$^{-1}$.
Although we detect PSRs B1822--14 and B0905--51 at high significance,
their bolometric luminosities are not anomalously high
($L_X\sim10^{31-32}$ erg s$^{-1}$) since they account for
$\sim$$10^{-3}$ of the pulsar spindown luminosity, consistent with the
values observed for many ordinary pulsars (see, e.g. Kargaltsev et
al.~2012, and references therein). Middle-aged and old pulsars show
thermal emission with $kT\approx0.1-0.3$ keV and small emission radii,
indicative of heating of the polar cap regions by a return current
from the pulsar magnetosphere. Based on this, to compute upper limits
on the bolometric luminosity for the non-detections, we consider a
thermal spectrum with $kT=0.2$ keV and the expected $N_{\rm H}$
computed using the empirical relation between the pulsar dispersion
measure and $N_{\rm H}$ found by \citet{He13}.  The low implied
luminositites indicate that all eight pulsars are genuinely middle
aged/old and are simply superposed on the SNRs by chance.

\section{Discussion and Conclusions}
To date, targeted deep searches have found no radio emission
associated with the three CCOs with measured periodicities
\citep{Gaensler00,Camilo04,Halpern07}.  It is possible that (for
as-yet-unknown reasons) the particle acceleration mechanism in these
objects never activates, rendering them permanently radio-quiet. As a
result, as they age and cool, these objects may eventually fade away
becoming undetectable in X-rays as well.  However, the sample of CCOs
is not yet large enough to know if they are intrinsically radio-quiet
as unfavorable viewing geometries may cause the radio beams to not
sweep towards us.

Previously, \citet{Got13b} explored the possibility that isolated
radio pulsars with $P > 20$ ms and $B_{\rm surf} < 3 \times 10^{10}$ G
are actually ``orphan'' CCOs, namely, neutron stars whose SNRs have
dissipated. In this scenario, CCOs maintain their weak magnetic fields
but at some stage become radio-loud. None of the 13 objects considered
were detected in X-rays, suggesting that CCOs and radio-loud pulsars
with $\sim$$10^{10}$ G are disjoint classes of objects. The dearth of
low-field, non-recycled radio pulsars in the portion of the
$P-\dot{P}$ diagram occupied by CCOs (Figure 1), despite no
observational selection effects against them, provides an additional
argument against this evolutionary outcome.

There is growing evidence that the apparently low magnetic fields of
CCOs ($\le10^{11}$ G, as inferred from $P$ and $\dot{P}$) may be the
result of a submerged field, which may eventually diffuse outward on
timescales of $\gtrsim$kyr \citep{Halpern10}. Therefore, CCOs may, in
principle, evolve into normal radio pulsars (with $B\gtrsim 10^{12}$
G) after $\gtrsim$1 kyr.  The buried field will diffuse back to the
surface on a time scale that is determined in large part by the amount
of mass accreted (Ho 2011; Vigan\'o \& Pons 2012).  Here, we have
investigated this possible evolutionary path of CCOs, assuming that
the neutron stars activate as radio pulsars in the process.  For the
first $\sim$$10^5$ yr, rapid field growth would move a CCO upward in
the $P-\dot{P}$ diagram.  The location of several pulsars from Table 1
in the $P-\dot{P}$ diagram directly above the CCO 1E 1207.4−-5209 is
consistent with this scenario.

Since there are 9 CCOs \citep{Halpern10} with ages
  $\lesssim$$10^4$ yr and $\sim$15 radio pulsars in SNRs with comparable
  ages \citep[see, e.g.,][and references therein]{Popov12}, if the CCO
  formation channel supplies a portion of the radio pulsar population,
  we estimate that $\sim$3 of the eight objects we have targeted
  should be CCO descendants.  However, none of the pulsars from the
volume-complete sample within 6 kpc exhibit unusually high X-ray
luminosity, implying that they are all genuinely middle aged/old
rotation-powered pulsars. This suggests that CCOs may not be latent
radio pulsars but may instead be permanently radio silent.
Alternatively, this could mean that the buried strong magnetic fields
in CCOs re-emerge long after they leave their birth site and/or the
SNR dissipates. If in the process they activate as radio pulsars, CCOs
may be hidden among the population of ordinary radio pulsars. In this
case, a systematic sensitive X-ray survey of isolated radio pulsars is
required to identify the ``orphan'' CCOs among them, using the
unusually high thermal X-ray luminosity as a discriminant.  If,
however, orphan CCOs remain radio-quiet, they would be difficult to
identify as X-ray pulsars unless they are relatively nearby.  As
suggested in \citet{Halpern13}, a possible candidate for a CCO
descendant is the nearby radio-quiet X-ray pulsar Calvera (1RXS
J141256.0+792204), although further investigation is necessary to
establish whether the apparent lack of radio and $\gamma$-ray emission
is simply due to an unfavorable viewing geometry.

Given that CCOs may account for up to $\sim$$1/3$ of neutron star
births \citep{Popov12}, if they represent a truly distinct population
the discrepancy between the Galactic core-collapse supernova rate
\citep[][]{Diehl06} and the neutron star formation rate is exacerbated
even further.

\acknowledgements 
We thank J.~P.~Halpern for supplying Figure 1. The
work presented was funded in part by NASA Astrophysics Data Analysis
Program (ADAP) grant NNX12AE24G awarded through Columbia
University. V.K. acknowledges funding from an NSERC Discovery Grant and
Accelerator Supplement, the Canadian Institute for Advanced Study,
FQRNT via le Centre de Recherche en Astrophysique du Quebec, the
Canada Research Chairs program, and the Lorne Trottier Chair in
Astrophysics and Cosmology.  A portion of the results presented was
based on observations obtained with \textit{XMM-Newton}, an ESA
science mission with instruments and contributions directly funded by
ESA Member States and NASA. This research has made use of the NASA
Astrophysics Data System (ADS), data obtained from the High Energy
Astrophysics Science Archive Research Center (HEASARC), provided by
NASA's Goddard Space Flight Center, and software provided by the
Chandra X-ray Center (CXC) in the application package CIAO.

Facilities: CXO (ACIS),XMM (EPIC)

\end{document}